\pdfoutput=1

\documentclass[12pt,a4paper]{article}
\def\tvalue{$T$-value\xspace}
\def\tvalues{$T$-values\xspace}
\def\tnull{$T_0$\xspace}
\def\pvalue{$p$-value\xspace}
\def\pvalues{$p$-values\xspace}

\usepackage{rotating}
\usepackage{pdflscape}

\usepackage{ifthen} 
\newboolean{pdflatex}
\setboolean{pdflatex}{true} 

\newboolean{articletitles}
\setboolean{articletitles}{true} 

\newboolean{uprightparticles}
\setboolean{uprightparticles}{false} 

\newboolean{inbibliography}
\setboolean{inbibliography}{false} 

\def\paperauthors{W. Barter, C. Burr, C. Parkes} 
\def\paperasciititle{Calculating p-values and their significances with the Energy Test for large datasets} 
\def\papertitle{Calculating $p$-values and their significances with the Energy Test for large datasets} 
\def\paperkeywords{{High Energy Physics}} 
\def\papercopyright{University of Manchester}

\def\paperlicenceurl{https://creativecommons.org/licenses/by/4.0/}

\usepackage[top=1in, bottom=1.25in, left=1in, right=1in]{geometry}

\usepackage{microtype}
\usepackage{lineno}  
\usepackage{xspace} 
\usepackage{caption} 

\usepackage{graphicx}  
\usepackage{color}
\usepackage{colortbl}
\graphicspath{{./figs/}} 

\usepackage{amsmath} 
\usepackage{amssymb}
\usepackage{amsfonts}
\usepackage{upgreek} 

\newcommand*\patchAmsMathEnvironmentForLineno[1]{%
\expandafter\let\csname old#1\expandafter\endcsname\csname #1\endcsname
\expandafter\let\csname oldend#1\expandafter\endcsname\csname
end#1\endcsname
 \renewenvironment{#1}%
   {\linenomath\csname old#1\endcsname}%
   {\csname oldend#1\endcsname\endlinenomath}%
}
\newcommand*\patchBothAmsMathEnvironmentsForLineno[1]{%
  \patchAmsMathEnvironmentForLineno{#1}%
  \patchAmsMathEnvironmentForLineno{#1*}%
}
\AtBeginDocument{%
\patchBothAmsMathEnvironmentsForLineno{equation}%
\patchBothAmsMathEnvironmentsForLineno{align}%
\patchBothAmsMathEnvironmentsForLineno{flalign}%
\patchBothAmsMathEnvironmentsForLineno{alignat}%
\patchBothAmsMathEnvironmentsForLineno{gather}%
\patchBothAmsMathEnvironmentsForLineno{multline}%
\patchBothAmsMathEnvironmentsForLineno{eqnarray}%
}

\usepackage{hyperxmp}

\usepackage[pdftex,
            pdfauthor={\paperauthors},
            pdftitle={\paperasciititle},
            pdfkeywords={\paperkeywords},
            pdfcopyright={Copyright (C) \papercopyright},
            pdflicenseurl={\paperlicenceurl}]{hyperref}

\usepackage[all]{hypcap} 

\usepackage{xspace} 
\usepackage{upgreek}

\def\MagUp {\mbox{\em Mag\kern -0.05em Up}\xspace}

\ifthenelse{\boolean{uprightparticles}}%
{

 \def\PDelta      {\ensuremath{\Delta}\xspace}                 
 \def\PXi      {\ensuremath{\Xi}\xspace}                 
 \def\PLambda      {\ensuremath{\Lambda}\xspace}                 
 \def\PSigma      {\ensuremath{\Sigma}\xspace}                 
 \def\POmega      {\ensuremath{\Omega}\xspace}                 
 \def\PUpsilon      {\ensuremath{\Upsilon}\xspace}                 
 
 \def\PB      {\ensuremath{\mathrm{B}}\xspace}                 
                  
 \def\PD      {\ensuremath{\mathrm{D}}\xspace}

 \def\PK      {\ensuremath{\mathrm{K}}\xspace}

 \def\Pi      {\ensuremath{\mathrm{i}}\xspace}

}
{

 \mathchardef\PDelta="7101
 \mathchardef\PXi="7104
 \mathchardef\PLambda="7103
 \mathchardef\PSigma="7106
 \mathchardef\POmega="710A
 \mathchardef\PUpsilon="7107
                  
 \def\PB      {\ensuremath{B}\xspace}                 
                  
 \def\PD      {\ensuremath{D}\xspace}

 \def\PK      {\ensuremath{K}\xspace}

 \def\Pi      {\ensuremath{i}\xspace}

}

\makeatletter
\ifcase \@ptsize \relax
  \newcommand{\miniscule}{\@setfontsize\miniscule{4}{5}}
\or
  \newcommand{\miniscule}{\@setfontsize\miniscule{5}{6}}
\or
  \newcommand{\miniscule}{\@setfontsize\miniscule{5}{6}}
\fi
\makeatother

\DeclareRobustCommand{\optbar}[1]{\shortstack{{\miniscule (\rule[.5ex]{1.25em}{.18mm})}
  \\ [-.7ex] $#1$}}

  \def\Kbar    {{\kern 0.2em\overline{\kern -0.2em \PK}{}}\xspace}

\def\KorKbar    {\kern 0.18em\optbar{\kern -0.18em K}{}\xspace}

  \def\Dbar    {{\kern 0.2em\overline{\kern -0.2em \PD}{}}\xspace}

\def\DorDbar    {\kern 0.18em\optbar{\kern -0.18em D}{}\xspace}

\def\Bbar    {{\ensuremath{\kern 0.18em\overline{\kern -0.18em \PB}{}}}\xspace}

\def\BorBbar    {\kern 0.18em\optbar{\kern -0.18em B}{}\xspace}

  \def\Y#1S{\ensuremath{\PUpsilon{(#1S)}}\xspace}

\def\Lbar        {{\ensuremath{\kern 0.1em\overline{\kern -0.1em\PLambda}}}\xspace}
\def\LorLbar    {\kern 0.18em\optbar{\kern -0.18em \PLambda}{}\xspace}

\def\AT#1     {\ensuremath{A_{\mathrm{T}}^{#1}}\xspace}           

\def\C#1      {\ensuremath{\mathcal{C}_{#1}}\xspace}                       
\def\Cp#1     {\ensuremath{\mathcal{C}_{#1}^{'}}\xspace}                    
\def\Ceff#1   {\ensuremath{\mathcal{C}_{#1}^{\mathrm{(eff)}}}\xspace}        
\def\Cpeff#1  {\ensuremath{\mathcal{C}_{#1}^{'\mathrm{(eff)}}}\xspace}       
\def\Ope#1    {\ensuremath{\mathcal{O}_{#1}}\xspace}                       
\def\Opep#1   {\ensuremath{\mathcal{O}_{#1}^{'}}\xspace}                    

\newcommand{\tev}{\ifthenelse{\boolean{inbibliography}}{\ensuremath{~T\kern -0.05em eV}}{\ensuremath{\mathrm{\,Te\kern -0.1em V}}}\xspace}
\newcommand{\gev}{\ensuremath{\mathrm{\,Ge\kern -0.1em V}}\xspace}
\newcommand{\mev}{\ensuremath{\mathrm{\,Me\kern -0.1em V}}\xspace}
\newcommand{\kev}{\ensuremath{\mathrm{\,ke\kern -0.1em V}}\xspace}
\newcommand{\ev}{\ensuremath{\mathrm{\,e\kern -0.1em V}}\xspace}
\newcommand{\gevc}{\ensuremath{{\mathrm{\,Ge\kern -0.1em V\!/}c}}\xspace}
\newcommand{\mevc}{\ensuremath{{\mathrm{\,Me\kern -0.1em V\!/}c}}\xspace}
\newcommand{\gevcc}{\ensuremath{{\mathrm{\,Ge\kern -0.1em V\!/}c^2}}\xspace}
\newcommand{\gevgevcccc}{\ensuremath{{\mathrm{\,Ge\kern -0.1em V^2\!/}c^4}}\xspace}
\newcommand{\mevcc}{\ensuremath{{\mathrm{\,Me\kern -0.1em V\!/}c^2}}\xspace}

\def\gsim{{~\raise.15em\hbox{$>$}\kern-.85em
          \lower.35em\hbox{$\sim$}~}\xspace}
\def\lsim{{~\raise.15em\hbox{$<$}\kern-.85em
          \lower.35em\hbox{$\sim$}~}\xspace}

\def\tell1  {TELL1\xspace}
\def\ukl1   {UKL1\xspace}

\usepackage{cite} 
\usepackage{mciteplus}

\usepackage{longtable}
\begin{document}

\renewcommand{\thefootnote}{\fnsymbol{footnote}}
\setcounter{footnote}{1}


\begin{titlepage}
\pagenumbering{roman}

\noindent
\begin{tabular*}{\linewidth}{lc@{\extracolsep{\fill}}r@{\extracolsep{0pt}}}
\ifthenelse{\boolean{pdflatex}}
 & & \today \\ 
 & & \\
\end{tabular*}

\vspace*{2.5cm}

{\normalfont\bfseries\boldmath\huge
\begin{center}
  \papertitle 
\end{center}
}

\vspace*{1.5cm}

\begin{center}
W. Barter, C. Burr, C. Parkes\\\vspace{1cm}
School of Physics and Astronomy, University of Manchester,\\
Oxford Road, Manchester, M13 9PL, UK\\
\end{center}

\vspace{\fill}

\begin{abstract}
  \noindent The energy test method is a multi-dimensional test of whether two samples are consistent with arising from the same underlying population, through the calculation of a single test statistic (called the \tvalue). The method has recently been used in particle physics to search for differences between samples that arise from CP violation. The generalised extreme value function has previously been used to describe the distribution of \tvalues under the null hypothesis that the two samples are drawn from the same underlying population. We show that, in a simple test case, the distribution is not sufficiently well described by the generalised extreme value function. We present a new method, where the distribution of \tvalues under the null hypothesis when comparing two large samples can be found by scaling the distribution found when comparing small samples drawn from the same population. This method can then be used to quickly calculate the \pvalues associated with the results of the test.
\end{abstract}

\vspace*{2.0cm}

\begin{center}
$ $
\end{center}

\vspace{\fill}
%

\end{titlepage}


\newpage
\setcounter{page}{2}
\mbox{~}


\renewcommand{\thefootnote}{\arabic{footnote}}
\setcounter{footnote}{0}



\pagestyle{plain} 
\setcounter{page}{1}
\pagenumbering{arabic}

\section{Introduction}
A key problem in data science is determining if two samples, measured in multi-dimensional spaces, are consistent with having arisen from the same underlying population. This can be alternatively phrased as asking whether there exists evidence for differences between two samples. One area where this question is crucial is in searches for direct CP violation, which can appear as local differences of event yields in the multi-dimensional phase space between samples of matter and anti-matter data. Several different approaches have been suggested to address this question. One approach that has recently been developed and used in the context of CP violation is that of the Energy Test Method~\cite{doi:10.1080/00949650410001661440,Aslan2005626,Williams:2011cd,Parkes:2016yie,LHCb-PAPER-2014-054,LHCb-PAPER-2016-044}. This method calculates a single test statistic, the energy or \tvalue, from the distribution of the data in the two samples. The value of this quantity can then be used to determine the consistency of the two samples by comparing to the expected distribution of the \tvalue under the null hypothesis that the two samples are drawn from the same population. We will label this the \tnull distribution. However, the analytic form that this distribution takes is not clearly understood. Understanding this distribution better will enable further application of the energy test method in the coming years, in samples containing a large number of events. 

Within the energy test method the \tvalue is calculated as
\begin{equation}
T =  \frac{1}{2}\frac{1}{n(n-1)}\sum_{i \neq j}^{n} \psi_{ij} + \frac{1}{2}\frac{1}{\overline{n}(\overline{n}-1)}\sum_{i \neq j}^{\overline{n}} \psi_{ij} - \frac{1}{n\overline{n}}\sum_{i, j}^{n, \overline{n}} \psi_{ij},
\label{eq:tvalue}
\end{equation}
where the first sum is over pairs of points (or events) in the first sample, the second sum is over pairs of points in the second sample, and the final sum is over both samples. There are $n$ events in the first sample and $\overline{n}$ events in the second sample. The value $\psi$ is a weighting function, commonly taken as a Gaussian of some scaled Euclidean distance squared ($d^2$) between two points in the (potentially multidimensional) phase space,
\begin{math}
\psi_{ij} = e^{-d^2/(2\delta^2)},
\end{math}
where $\delta$ is a length-scale that is optimised for the problem under consideration. However, other choices have been made for this weighting function (for example logarithms of the Euclidean distance between points). The energy test essentially calculates the mean of the $\psi$ distribution using pairs of events taken from each sample independently (the first two terms), and then considers cross-terms between the two samples (the final term), with the added consideration that since the same events are used many times to calculate distances, the different terms in the sums that calculate these means are correlated. In the case where the two samples are drawn from the same population the expected \tvalue is 0. A large, positive \tvalue indicates differences between the samples. The method has recently been extended to also enable comparison of samples containing impurities~\cite{Parkes:2016yie}.

Estimating the \tnull distribution is an important open question that we consider in this article. This question lies at the heart of interpreting whether the returned test statistic is significant, and whether the two samples are consistent with being drawn from the same underlying population. To date, the typical approach to find this distribution when analysing data is to randomly assign the data to be tested to two samples, and calculate the \tvalue of these permuted samples, which are created, by definition, under the null hypothesis (taking a `permutation approach'). By repeating this process multiple times the \tnull distribution of \tvalues under the null hypothesis can be found, and used to determine the \pvalue associated with the nominal test, by counting how often the permutations return a more extreme \tvalue than the true data. If the behaviour of the \tnull distribution is not known, then one may need to repeat this process over 1.7 million times to determine whether a \tvalue has a probability (\pvalue) of less than $6\times 10^{-7}$ (i.e. in the case where data returns a large, positive \tvalue, and clearly lies in the tail of the \tnull distribution). Such a probability is equivalent to that of finding a result 5 Gaussian standard deviations (with $5\sigma$ significance) from an expected central value, and is the key criteria in particle physics for announcing a discovery.

The use of this permutation approach raises a problem when considering large samples. The calculation of the \tvalue is computationally intensive, and scales as $\mathcal{O}(n^2)$, where $n$ is the number of events in the sample (since $\mathcal{O}(n^2)$ distances must be calculated, used to determine some weighting function, and then summed). Similar calculations must be run many times to accurately probe the tail of the \tnull distribution, since the tail must be understood if a significant result is to be claimed. This increases the time taken in data analysis, and potentially makes the energy test significantly less useful as a data analysis method as sample sizes get larger and if a large number of permutations are required. A clear understanding of the \tnull distribution will therefore allow the use of the energy-test to analyse the largest data samples, making its use tractable in cases where it was not before.

It has been suggested~\cite{Aslan2005626} that the \tnull distribution seems to follow a generalised extreme value distribution when the energy test is used as a goodness of fit test.\footnote{The article that introduced this approach~\cite{Aslan2005626} noted that no proof exists that the GEV function describes the distribution in question and that the behaviour must be verified for each specific case.} However, this property has subsequently been used when the energy test is used as a two sample test~\cite{Williams:2011cd,Parkes:2016yie,LHCb-PAPER-2014-054,LHCb-PAPER-2016-044}, despite no suggestion that the \tnull distribution takes this form in the initial paper~\cite{Aslan2005626}. The assumption of this property has allowed quick calculations of \pvalues and significances, since a generalised extreme value (GEV) function can be fit to the \tnull distribution found from a limited number of permuted samples, and used to determine the significance of the \tvalue obtained when analysing the two samples present in the data. This is particularly useful if the sample sizes are too large to generate a large number of permutations quickly, since it allows large significances to be inferred when the \tvalue in data lies in the tail of the distribution, and is potentially larger than any of the \tvalues found in the permuted samples. This method has been used to determine \pvalues, often alongside direct counts of how often the permuted samples return more extreme \tvalues than the true data~\cite{Williams:2011cd,Parkes:2016yie,LHCb-PAPER-2014-054,LHCb-PAPER-2016-044}. With more studies using the energy test expected in the future, we investigate here whether the GEV function adequately describes the tail of the \tnull distribution in a test case of the two sample comparison problem, and address this question of how to model the distribution of \tvalues found under the null hypothesis: key for quickly performing the energy test and finding an unbiased and precise \pvalue if the sample \tvalue is located in the tail of the \tnull distribution. Formal mathematical proofs are not presented in this article. Instead, two different models are considered and used in toy studies to demonstrate the shortcomings and power of different methods. These models are presented in the next section. Following this, the use of the GEV function and a new approach are both presented.

\section{Data samples}
Two different toy models are considered for these studies:
\begin{enumerate}
\item Model 1\\ 
This model is a very simple model, where events in each sample are generated according to a uniform distribution in three dimensions, with the allowed region of phase space being located between 0 and 1 in all dimensions. Here the probability density of a point in phase space is independent of the location in the space.  
\item Model 2\\ 
We also use the more complicated, physically-motivated model, first set out in Ref.~\cite{Williams:2010vh}, and also studied in the context of the energy test in Ref.~\cite{Williams:2011cd} and Ref.~\cite{Parkes:2016yie}. This model considers the decay of a particle X to a three-body final state. The axes considered in this problem are formed from the three different invariant mass combinations of pairs of the final state particles. The presence of intermediate resonances mean that the density of events depends on the location in the phase space. This model is generated using the Laura++ package~\cite{Laura}.
\end{enumerate}
In both cases two different choices of the function $\psi_{ij}$ are used: a Gaussian, as set out above, and a logarithmic function $-\log|d+\epsilon|$, where $\epsilon$, like $\delta$ above, is a parameter to be chosen by the analyst, and which can be optimised for the study in question.

\section{Using the generalised extreme value function}
The generalised extreme value function takes three free parameters, $\mu$, $\sigma$ and $\xi$,\footnote{The literature also uses the parameter $c = -\xi$.} and describes the distribution of a variable, labelled here as $x$. For $\xi < 0$ the value is non-zero for $-\infty < x < \mu - \sigma/\xi$, while for $\xi > 0$ it is non-zero for $\mu - \sigma/\xi < x < +\infty$. In the special case where $\xi =0$ it is non-zero for all values of x on the real axis. The cumulative distribution is given by
\begin{equation}
\text{CDF}(x) = \exp\big(-\big(1 + \xi(\frac{x-\mu}{\sigma})\big)^{-1/\xi}\big),
\end{equation}
in the relevant ranges defined above, for $\xi\neq 0$. In the case where $\xi = 0$, the cumulative distribution is
\begin{equation}
\text{CDF}(x) = \exp(-\exp(-(x-\mu)/\sigma)).
\end{equation}

It is not obvious whether the GEV function should be used to describe the \tnull distribution. If we take a weighting function for the energy test that can only take values between 0 and 1, such as the Gaussian function discussed above, then the \tnull distribution must be contained in the range -1 to 1, given the form of equation~\ref{eq:tvalue}. However, any fit of the GEV function must contain either a finite probability for a \tvalue above 1 if $\xi \geq 0$, and/or below -1 if $\xi\leq 0$. Therefore, regardless of the value of $\xi$, with the GEV function the cumulative distribution is never solely defined in the range $-1<T<1$; it is therefore possible to find a \tvalue where the significance is either underestimated or overestimated.

\begin{figure}[tb]
  \begin{center}
    \includegraphics[width=0.49\linewidth]{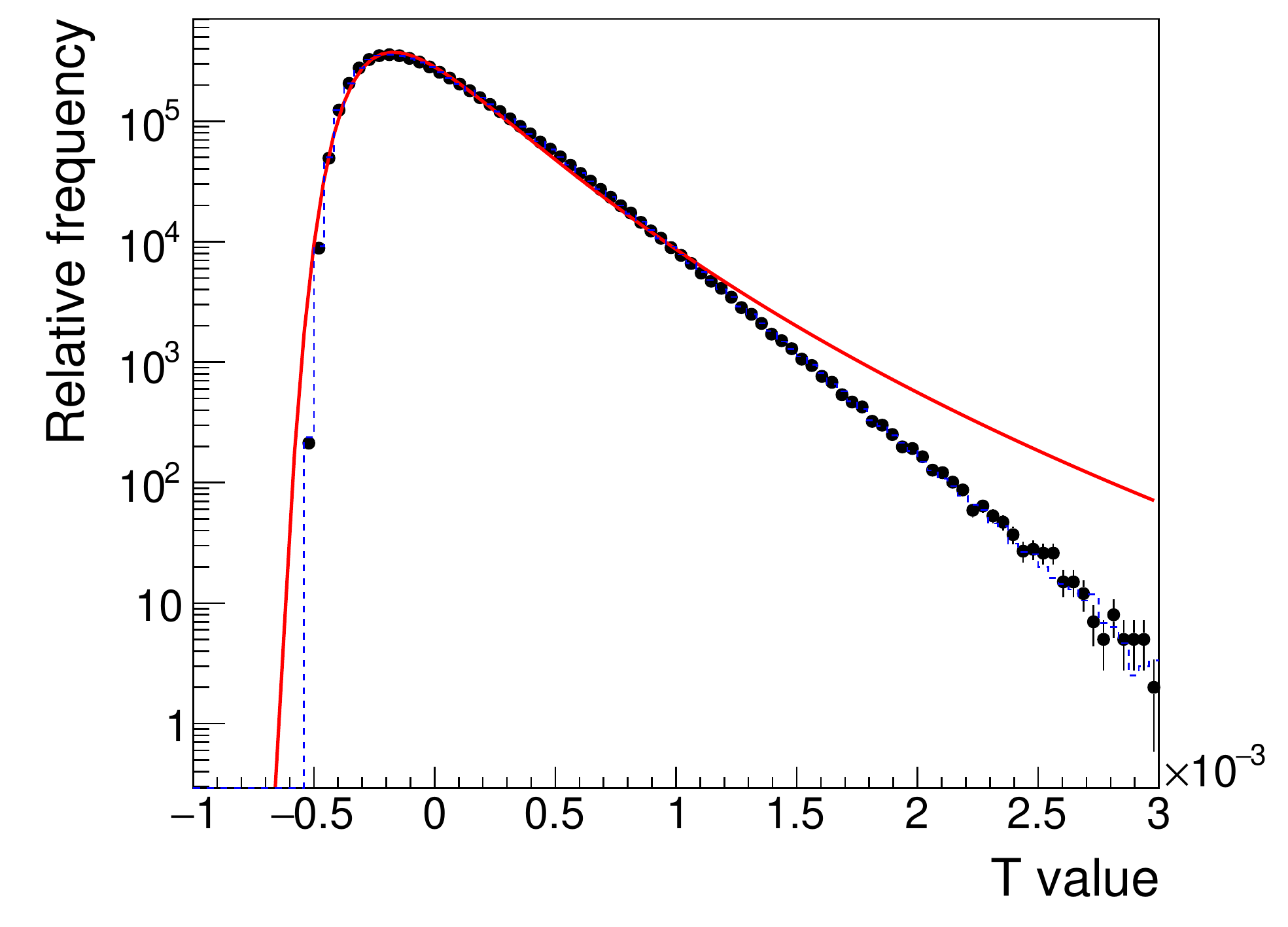}
    \includegraphics[width=0.49\linewidth]{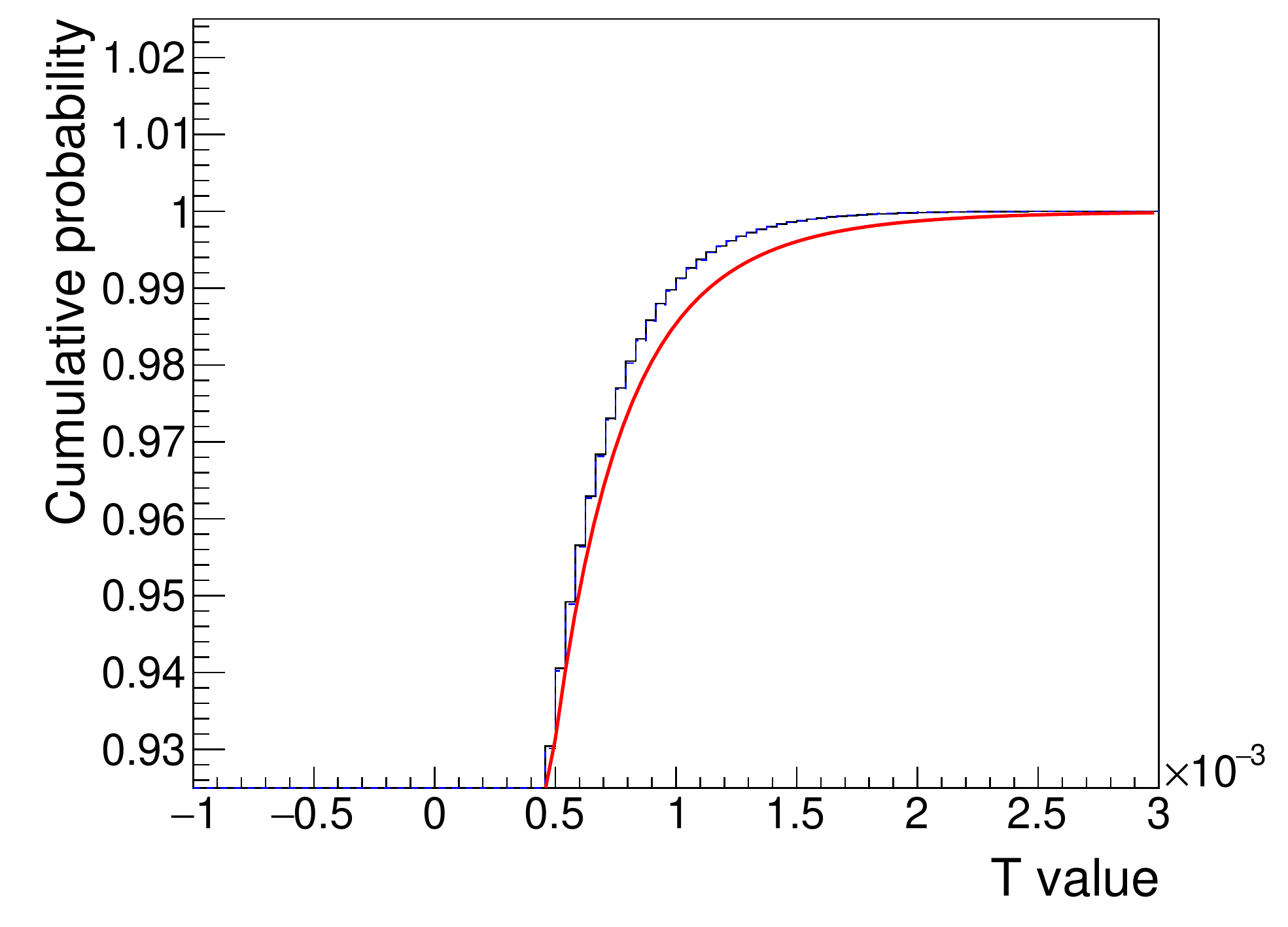}
    \vspace*{-0.5cm}
  \end{center}
  \caption{ 
    The (left) \tvalues found using the Gaussian weighting function are shown as black points (for comparing two 1000-event samples generated using model 1), with the fit of a GEV function shown as the solid red line; (right) the corresponding cumulative distribution. The dashed blue line corresponds to scaled \tvalues where only 500 events are used in each sample. On the cumulative plot the differences between the 1000 and 500 event \tnull distributions are too small to be visible.}
  \label{fig:Gaus_GEV}
\end{figure}

\begin{figure}[tb]
  \begin{center}
    \includegraphics[width=0.49\linewidth]{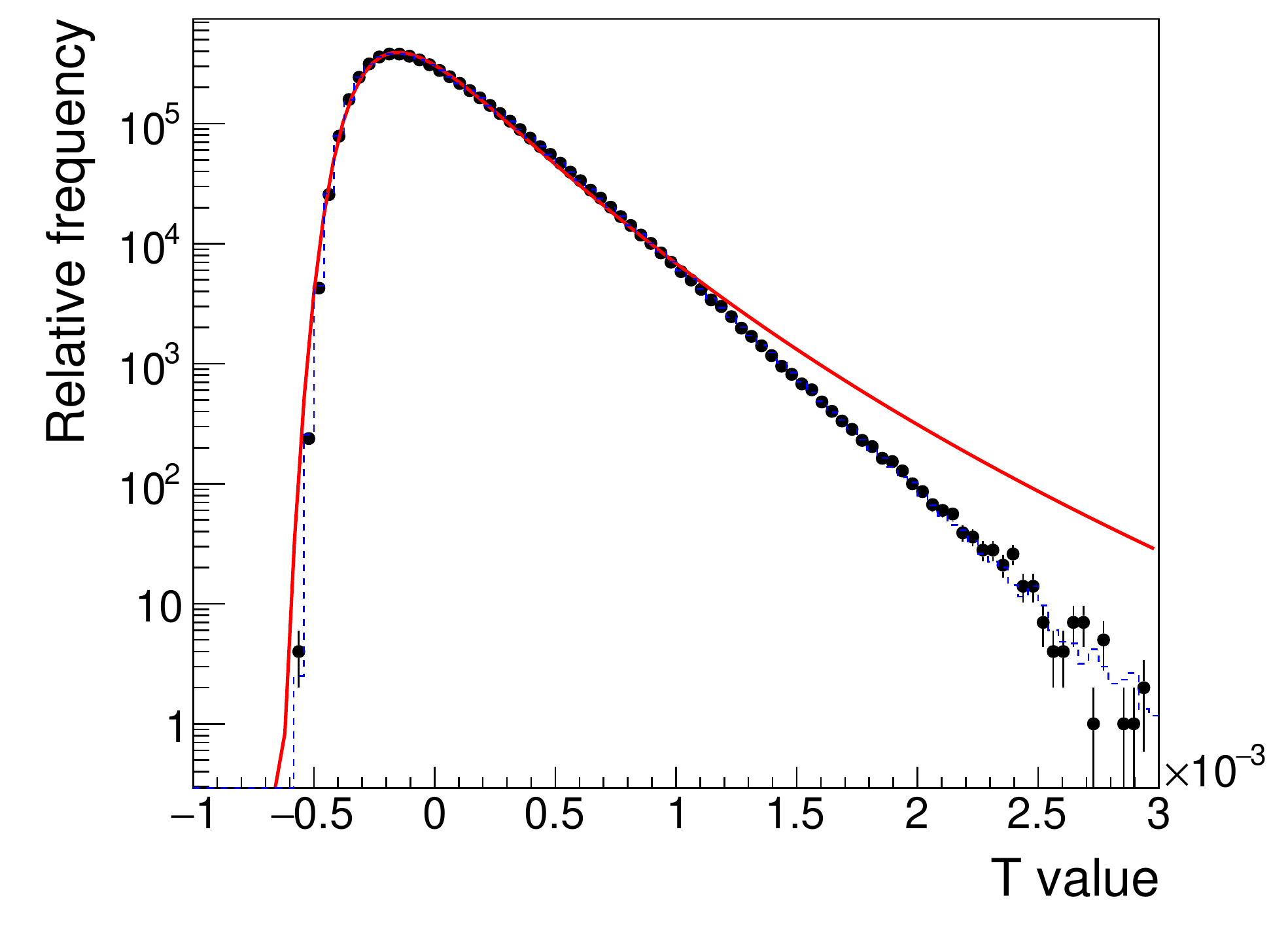}
    \includegraphics[width=0.49\linewidth]{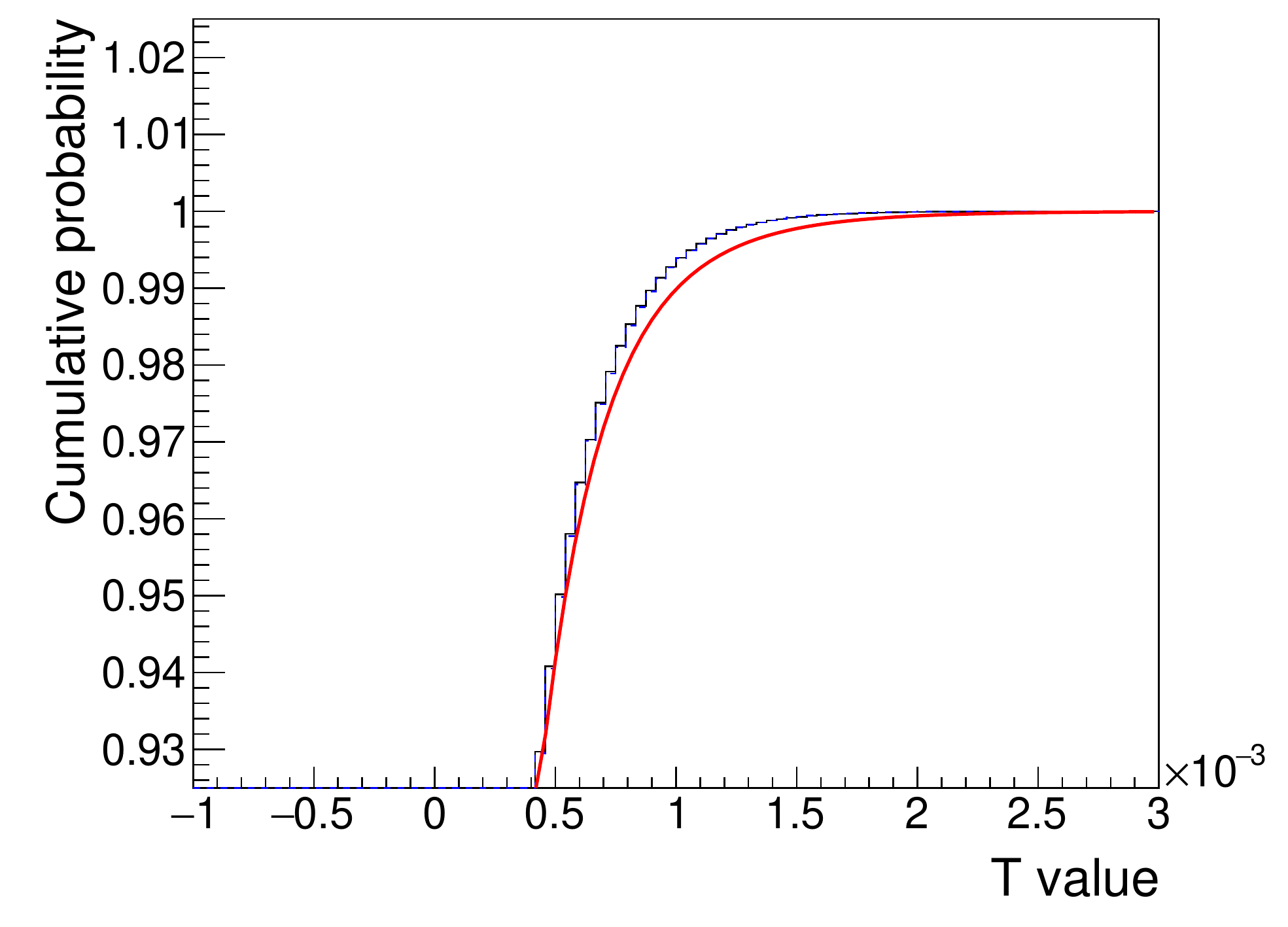}
    \vspace*{-0.5cm}
  \end{center}
  \caption{  
    The figure contains the same information as Figure~\ref{fig:Gaus_GEV} but using the logarithmic weighting function.
    }
  \label{fig:Log_GEV}
\end{figure}

We perform empirical studies to determine whether this poses problems when seeking to describe reasonable \tnull distributions that can be generated, or if the GEV function is never-the-less sufficient to describe the distribution at a level needed within particle physics studies. We generate two samples in Model 1, each containing 1000 points (or events), and calculate the \tvalue. This is done twice: once for the Gaussian weighting function and once for the logarithmic function. This is repeated 5~million times in order to create a distribution of \tvalues for each weighting function (a permutation approach is not used, since `toy' data can be quickly generated to find the distribution of \tvalues). With both samples generated from the same underlying distribution, the distributions of \tvalues found are the relevant \tnull distributions for each weighting function. For both the Gaussian weighting function and the logarithmic function we use $\delta=\epsilon=0.5$. The \tnull distributions are shown in Figures \ref{fig:Gaus_GEV} and \ref{fig:Log_GEV}. A GEV function is fit to the distributions in both cases using the binned maximum likelihood method, with the fit also shown on the same figures, alongside the cumulative distribution and the cumulative distribution associated with the best fit. In these fits a normalisation, and the three parameters ($\mu,\sigma,\xi$), are left unconstrained. It is clear for both weighting functions that the fit function does not describe the data. In addition, further cross-checks are made to see if the GEV function can be used to describe the data. First, $\chi^2$ fits are performed, and no good $\chi^2$ value is returned. Second, the parameters in the GEV function left unconstrained in the fit are reduced to a normalisation and a value of $\xi$. This is achieved by fixing $\mu$ using the relation that the cumulative probability is $1/\text{e}$ when $x=\mu$, and that when the cumulative distribution is 0.5, the relevant $x$-value, $x_{1/2}$ (the median) can be used to relate $\sigma$ and $\xi$ through $\sigma = \xi (x_{1/2} - \mu)/((\log(2))^{-\xi} - 1)$. Once again, in this case no good fit is observed. As a cross-check, rather than generating new samples of data repeatedly, the permutation approach is taken to determine the \tnull distribution. This does not change our conclusions. In the simple model studied here the fit returns a value for the parameter $\xi > 0$, and we find that the p-value tends to be overestimated in the tail of the distribution at large $T$. In this case the discovery of a new effect might be missed. If the fit returns a value of $\xi < 0$, the p-value tends to be underestimated, and such a discovery may be incorrectly claimed.


This finding represents an important note in the application of the Energy Test method: the GEV function has already been used in the literature to fit the \tnull distribution, and to determine \pvalues by extrapolating the fit results. This toy model is a simple case where this method is not valid: in this case, the use of the GEV function leads to the over-estimation of the \pvalue for large \tvalues, and a new effect or discovery might be missed. We therefore turn our attention to a novel approach to efficiently determining the \tnull distribution.

\section{Scaling the T-values}
The energy test (as set out in Equation~1) corresponds to the calculation of the mean values of the weighting function $\psi$, using correlated inputs, with $\mathcal{O}(n^2)$ terms (or similar) in each sum. If all the terms in calculating these means were not correlated (i.e. the same events were not re-used to calculate multiple distances), then increasing the sample sizes by a common factor $k$ would simply scale the behaviour of each of the three terms in the \tnull distribution: the distribution of $kT$ would be independent of the sample size (for sufficiently large samples). We investigate empirically whether this property also holds here.

We first test using model 1, by repeating the previous studies, but using two samples of 500 events (as opposed to the 1000 event samples set out above). This is repeated 30 million times to find the \tnull distribution. The calculated \tvalues are scaled by a factor of $500/1000 = 0.5$, and overlaid on Figures~\ref{fig:Gaus_GEV} and~\ref{fig:Log_GEV}. This gives a much better description of the \tvalues in the 1000 event samples than the GEV function. 
 
\begin{figure}[tb]
  \begin{center}
  \includegraphics[width=0.495\linewidth]{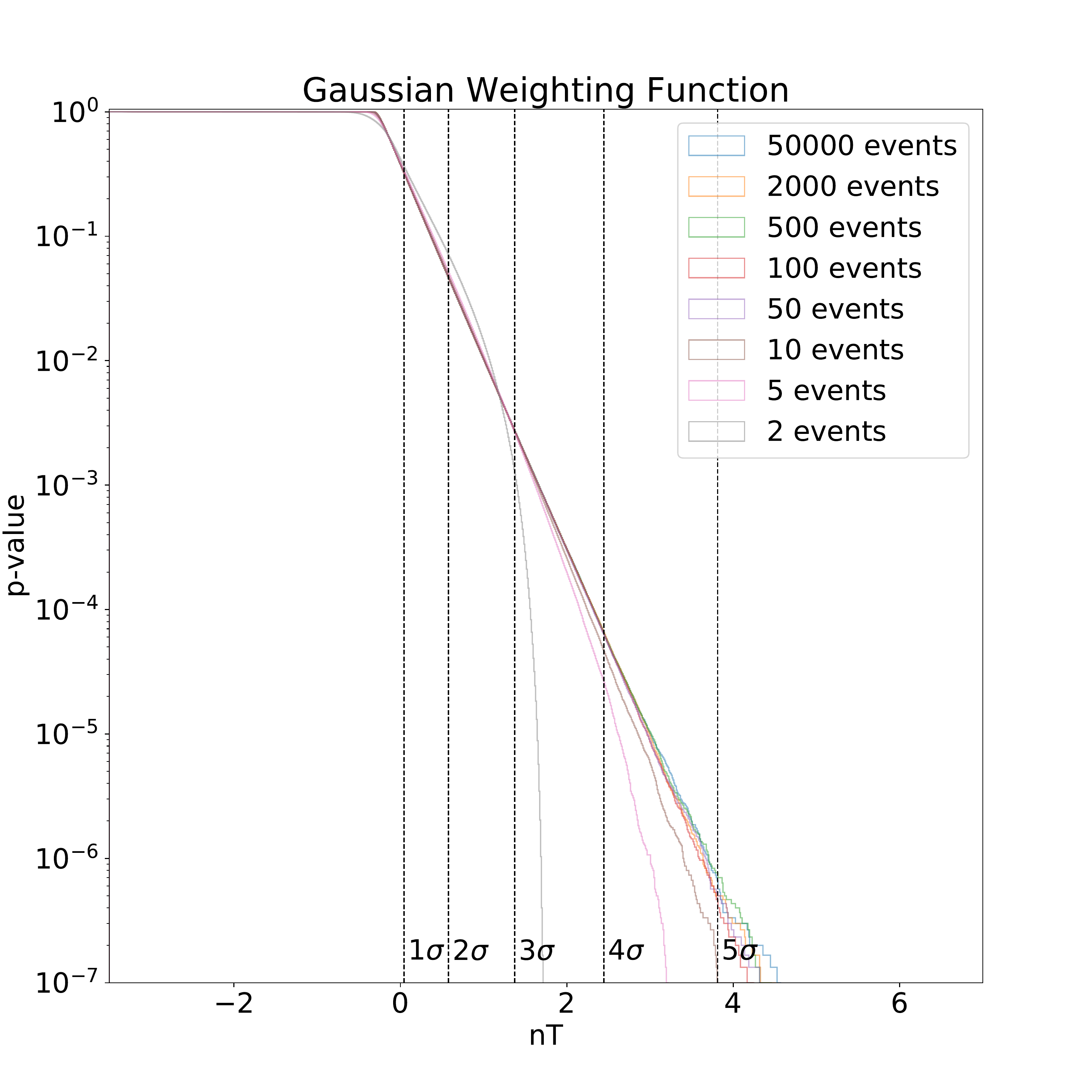}
    \includegraphics[width=0.495\linewidth]{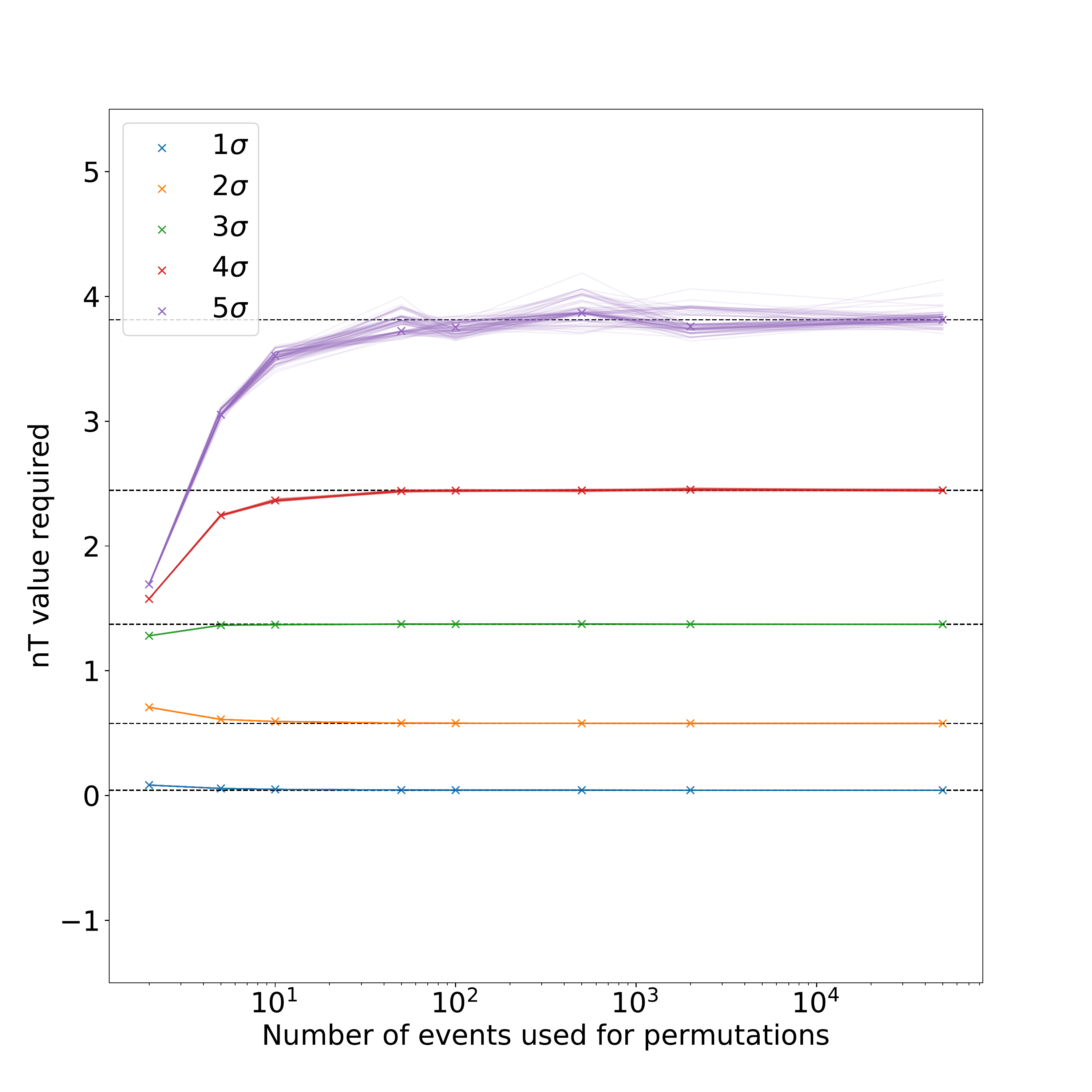}
    \vspace*{-0.5cm}
  \end{center}
  \caption{
    (Left) The \pvalues associated with a particular value of $nT$ in model 2, considering the Gaussian weighting function (also known as the survival function). This is shown for different values of $n$, with the \pvalues associated with 1, 2, 3, 4, and 5 $\sigma$ significance marked as vertical lines for the largest value of $n$ we consider, $n_\text{max}= 50\,000$. (Right) The values of $nT$ required for these \pvalues are plotted as a function of $n$. The uncertainty on these values is estimated and shown using the semi-transparent lines. These lines are generated by bootstrapping (sampling with replacement) a new set of \tvalues from the original set and determining the relevant value of $nT$ in this new sample. This is repeated 100 times. For ease of comparison between different values of $n$, the expected values of $nT$ required to achieve 1, 2, 3, 4, and 5 $\sigma$ significance for $n=n_\text{max}= 50\,000$ are also shown as horizontal lines.}
  \label{fig:Gaus_scale}
\end{figure}

\begin{figure}[tb]
  \begin{center}
\includegraphics[width=0.495\linewidth]{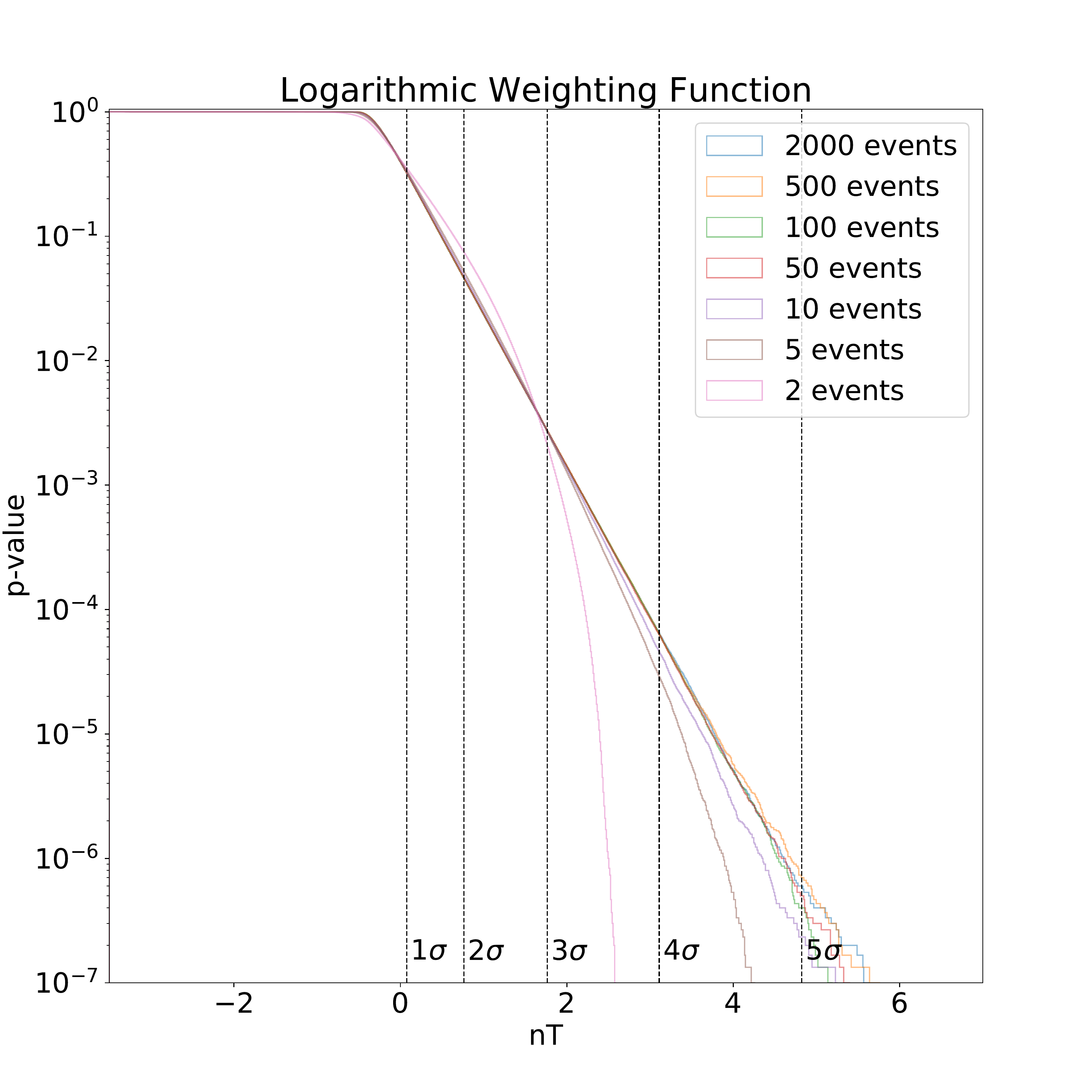}
    \includegraphics[width=0.495\linewidth]{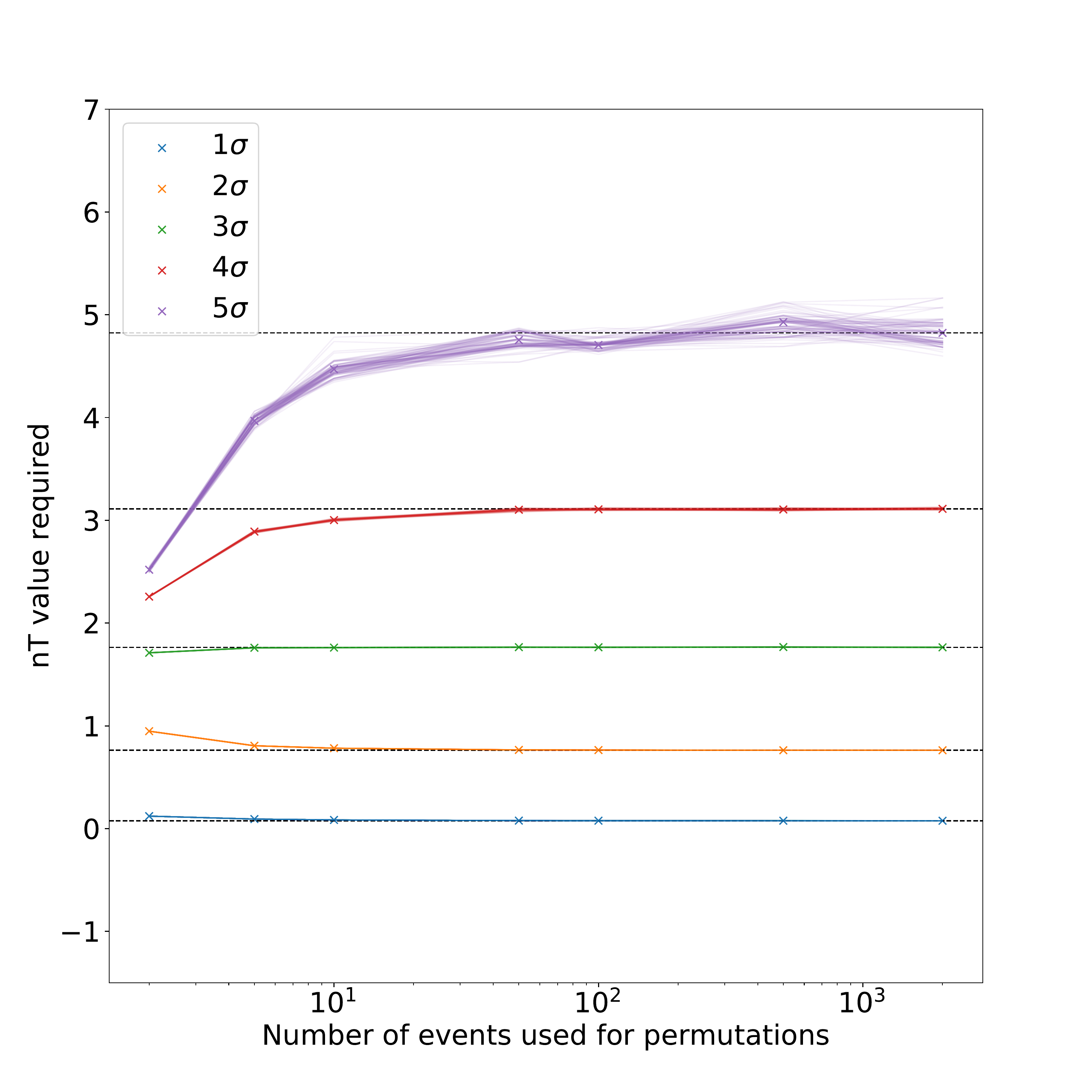}
    \vspace*{-0.5cm}
  \end{center}
  \caption{
    The figure contains similar information to Figure~\ref{fig:Gaus_scale} but using the logarithmic weighting function, and with $n_\text{max}= 2\,000$. The smaller value of $n_\text{max}$ is used here owing to the significant computing time taken to calculate 30 million \tvalues for large $n$.
    }
  \label{fig:Log_scale}
\end{figure}

We make further studies by generating a sample using model 2 containing $1\,000\,000$ events. We use this sample to investigate the \tnull distribution further: we randomly assign events to two smaller sub-samples (here taken to contain an equal number of events, $n$) and calculate the \tvalue and the value of $nT$ associated with running the energy test to compare these sub-samples. This is performed 30 million times for each value of $n$ considered. The distribution of $nT$ is shown in Figures~\ref{fig:Gaus_scale} and~\ref{fig:Log_scale} for different values of $n$, for the Gaussian and logarithmic weighting functions (in both cases we use  $\delta=\epsilon=0.5$). Also shown are the values of the $nT$ distribution where the \pvalue for getting such a result corresponds to 1, 2, 3, 4, and 5 $\sigma$ evidence for differences between the samples respectively, for different values of $n$.


It is clear that in these cases the value of $nT$ associated with a particular level of significance is independent of $n$ for sufficiently large n (typically around 100 events, for the levels of significance we have investigated). We have also applied this method to additional models, and used different weighting functions ($\psi$) and have found no evidence for the breaking of this scaling property. 
We also find that this scaling property does not rely on the sample sizes being equal, so that if the sizes of the samples being compared are increased by a factor $k$, the distribution of $kT$ under the null hypothesis is invariant (for large sample sizes).
However, we note that we have no formal proof of this property. Indeed, for the Gaussian weighting function, the distribution the value of $nT$ must lie in the range $-n < nT < n$. Therefore the use of this scaling property to estimate \pvalues will also provide incorrect coverage for some large (positive or negative) value of $T$. However, in the cases we have examined, this effect appears negligible for $n$ larger than about 100 points when considering significances of around $5\sigma$ and smaller. We recommend similar tests are performed for each specific case where the method is used.

This scaling property means that the \tnull distribution can be generated using a small value of n, and then scaled to determine the appropriate \tnull distribution for the sample sizes under consideration in the main test. This speeds up the computation of the significance of \tvalues when the \pvalue is small: with this method the calculation of the distribution of the null hypothesis no longer requires the generation of over one million permuted samples \emph{of the same size as the initial samples} to claim a \pvalue smaller than 5 Gaussian standard deviations. Instead, the \tnull distribution can be quickly generated using small samples; only the one calculation of the \tvalue of the main data sample remains computationally intensive, and uses the full event yield in the calculation. Consequently it is with the biggest datasets that the impact of this scaling property is most significant.

\section{Conclusions}
The energy test is a standard method within data science that measures whether two samples are consistent with arising from the same underlying population.
The method has recently been used for studies in particle physics. The small \pvalues necessary to claim a discovery in particle physics require understanding rare \tvalues returned by the energy test method under the null hypothesis that the two samples are identical. Problems arise if the datasets under study are so large that the distribution under the null hypothesis cannot be simulated quickly with a permutation method, so the tail of this distribution cannot be studied sufficiently. In the existing literature the generalised extreme value distribution has been fit to the distribution, with an extrapolation of the fit then used to determine the \pvalues associated with large \tvalues~\cite{Williams:2011cd,Parkes:2016yie,LHCb-PAPER-2014-054,LHCb-PAPER-2016-044} (often alongside a straightforward `counting method' of determining the \pvalue from how often tests of the permuted samples return \tvalues larger than that in the test of the true samples). However, we have shown here for a simple test case that the \tnull distribution is not sufficiently well described by the generalised extreme value function. We have therefore also presented a new method where small sub-samples of the data, which can be analysed quickly, can be used to find the distribution of \tvalues expected under the null hypothesis for large sample sizes. In this way, the tail of the distribution under the null hypothesis can be probed. This allows the accurate determination of small \pvalues associated with claims of discovery of new physical phenomena.

\section*{Acknowledgements}
We wish to thank Roger Barlow, Igor Babuschkin, Giulio Dujany, Marco Gersabeck, Gediminas Sarpis, Mike Williams, and G{\"u}nter Zech for illuminating discussions. We also thank G{\"u}nter Zech for the clarification that if the GEV function describes the distribution when used for goodness-of-fit studies (as considered in Ref.~\cite{Aslan2005626}) this does not imply it also describes the distribution for two sample tests, despite its subsequent use in such studies. 
This work was supported by STFC grant number ST/N000374/1.

\addcontentsline{toc}{section}{References}
\setboolean{inbibliography}{true}
\bibliographystyle{LHCb}
\bibliography{main,LHCb-PAPER}

\ifx\mcitethebibliography\mciteundefinedmacro
\PackageError{LHCb.bst}{mciteplus.sty has not been loaded}
{This bibstyle requires the use of the mciteplus package.}\fi
\providecommand{\href}[2]{#2}
\begin{mcitethebibliography}{1}
\mciteSetBstSublistMode{n}
\mciteSetBstMaxWidthForm{subitem}{\alph{mcitesubitemcount})}
\mciteSetBstSublistLabelBeginEnd{\mcitemaxwidthsubitemform\space}
{\relax}{\relax}

\bibitem{doi:10.1080/00949650410001661440}
B.~Aslan and G.~Zech, \ifthenelse{\boolean{articletitles}}{\emph{New test for
  the multivariate two-sample problem based on the concept of minimum energy},
  }{}\href{http://dx.doi.org/10.1080/00949650410001661440}{J.\ Stat.\ Comput.\
  Simul.\  \textbf{75} (2005) 109}\relax
\mciteBstWouldAddEndPuncttrue
\mciteSetBstMidEndSepPunct{\mcitedefaultmidpunct}
{\mcitedefaultendpunct}{\mcitedefaultseppunct}\relax
\EndOfBibitem
\bibitem{Aslan2005626}
B.~Aslan and G.~Zech, \ifthenelse{\boolean{articletitles}}{\emph{Statistical
  energy as a tool for binning-free, multivariate goodness-of -fit tests,
  two-sample comparison and unfolding},
  }{}\href{http://dx.doi.org/10.1016/j.nima.2004.08.071}{Nucl.\ Instrum.\
  Meth.\  \textbf{A537} (2005) 626 }\relax
\mciteBstWouldAddEndPuncttrue
\mciteSetBstMidEndSepPunct{\mcitedefaultmidpunct}
{\mcitedefaultendpunct}{\mcitedefaultseppunct}\relax
\EndOfBibitem
\bibitem{Williams:2011cd}
M.~Williams, \ifthenelse{\boolean{articletitles}}{\emph{{Observing \CP
  violation in many-body decays}},
  }{}\href{http://dx.doi.org/10.1103/PhysRevD.84.054015}{Phys.\ Rev.\
  \textbf{D84} (2011) 054015},
  \href{http://arxiv.org/abs/1105.5338}{{\normalfont\ttfamily
  arXiv:1105.5338}}\relax
\mciteBstWouldAddEndPuncttrue
\mciteSetBstMidEndSepPunct{\mcitedefaultmidpunct}
{\mcitedefaultendpunct}{\mcitedefaultseppunct}\relax
\EndOfBibitem
\bibitem{Parkes:2016yie}
C.~Parkes {\em et~al.}, \ifthenelse{\boolean{articletitles}}{\emph{{On
  model-independent searches for direct CP violation in multi-body decays}},
  }{}\href{http://dx.doi.org/10.1088/1361-6471/aa75a5}{J.\ Phys.\  \textbf{G44}
  (2017), no.~8 085001},
  \href{http://arxiv.org/abs/1612.04705}{{\normalfont\ttfamily
  arXiv:1612.04705}}\relax
\mciteBstWouldAddEndPuncttrue
\mciteSetBstMidEndSepPunct{\mcitedefaultmidpunct}
{\mcitedefaultendpunct}{\mcitedefaultseppunct}\relax
\EndOfBibitem
\bibitem{LHCb-PAPER-2014-054}
LHCb collaboration, R.~Aaij {\em et~al.},
  \ifthenelse{\boolean{articletitles}}{\emph{{Search for $\CP$ violation in
  $\Dz\to\pim\pip\piz$ decays with the energy test}},
  }{}\href{http://dx.doi.org/10.1016/j.physletb.2014.11.043}{Phys.\ Lett.\
  \textbf{B740} (2015) 158},
  \href{http://arxiv.org/abs/1410.4170}{{\normalfont\ttfamily
  arXiv:1410.4170}}\relax
\mciteBstWouldAddEndPuncttrue
\mciteSetBstMidEndSepPunct{\mcitedefaultmidpunct}
{\mcitedefaultendpunct}{\mcitedefaultseppunct}\relax
\EndOfBibitem
\bibitem{LHCb-PAPER-2016-044}
LHCb collaboration, R.~Aaij {\em et~al.},
  \ifthenelse{\boolean{articletitles}}{\emph{{Search for \CP violation in the
  phase space of $\Dz\to \pip\pim\pip\pim$ decays}},
  }{}\href{http://dx.doi.org/10.1016/j.physletb.2017.05.062}{Phys.\ Lett.\
  \textbf{B769} (2017) 345},
  \href{http://arxiv.org/abs/1612.03207}{{\normalfont\ttfamily
  arXiv:1612.03207}}\relax
\mciteBstWouldAddEndPuncttrue
\mciteSetBstMidEndSepPunct{\mcitedefaultmidpunct}
{\mcitedefaultendpunct}{\mcitedefaultseppunct}\relax
\EndOfBibitem
\bibitem{Williams:2010vh}
M.~Williams, \ifthenelse{\boolean{articletitles}}{\emph{{How good are your
  fits? Unbinned multivariate goodness-of-fit tests in high energy physics}},
  }{}\href{http://dx.doi.org/10.1088/1748-0221/5/09/P09004}{JINST \textbf{5}
  (2010) P09004}, \href{http://arxiv.org/abs/1006.3019}{{\normalfont\ttfamily
  arXiv:1006.3019}}\relax
\mciteBstWouldAddEndPuncttrue
\mciteSetBstMidEndSepPunct{\mcitedefaultmidpunct}
{\mcitedefaultendpunct}{\mcitedefaultseppunct}\relax
\EndOfBibitem
\bibitem{Laura}
J.~Back {\em et~al.}, \ifthenelse{\boolean{articletitles}}{\emph{{Laura++ : a
  Dalitz plot fitter}},
  }{}\href{http://arxiv.org/abs/1711.09854}{{\normalfont\ttfamily
  arXiv:1711.09854}}\relax
\mciteBstWouldAddEndPuncttrue
\mciteSetBstMidEndSepPunct{\mcitedefaultmidpunct}
{\mcitedefaultendpunct}{\mcitedefaultseppunct}\relax
\EndOfBibitem
\end{mcitethebibliography}
 
\end{document}